\documentclass[prd, twocolumn, superscriptaddress, nofootinbib]{revtex4}
\usepackage{epsfig}

\newcommand{\beq}{\begin{equation}}
\newcommand{\eeq}{\end{equation}}
\newcommand{\beqa}{\begin{eqnarray}}
\newcommand{\eeqa}{\end{eqnarray}}
\newcommand{\om}{\Omega_m}

\newcommand{\dl}{\Delta}

\begin{document} 
\title{Dark Energy, Expansion History of the Universe, and SNAP} 
\author{Eric V. Linder}
\affiliation{Physics Division, Berkeley Lab, Berkeley, CA 94720} 

\begin{abstract} 
This talk presents a pedagogical discussion of how precision 
distance-redshift observations can map out 
the recent expansion history of the universe, including the present 
acceleration and the transition to matter dominated deceleration. 
The proposed Supernova/Acceleration Probe (SNAP) will carry out 
observations determining 
the components and equations of state of the energy density, 
providing insights into the cosmological model, the nature of 
the accelerating dark energy, and potentially clues to fundamental 
high energy physics theories and gravitation.  This includes the 
ability to distinguish between various 
dynamical scalar field models for the dark energy, as well as 
higher dimension and alternate gravity theories.  
A new, advantageous parametrization for the study of 
dark energy to high redshift is also presented. 
\end{abstract} 

\maketitle 

\section{Introduction}\label{intro} 

Little else evokes the recent great advances in our abilities in 
cosmological observations like the quest to explore the expansion 
history of the universe.  This carries cosmology well beyond 
``determining two numbers'' -- the present 
dimensionless density of matter $\Omega_m$ and the present 
deceleration parameter $q_0$ of Sandage \cite{san61} -- to 
seeking to reconstruct the entire 
function $a(t)$ representing the expansion history of the 
universe.  In the previous case cosmologists sought only 
a local measure -- 
the first two derivatives of the scale factor $a$, evaluated at 
a single time $t_0$ -- while in the latter case we strive to map 
out the function determining the global dynamics 
of the universe. 

While many qualitative elements of cosmology follow merely from the 
form of the metric, i.e.~the kinematical cosmology (see Weinberg 
\cite{wein}), 
deeper understanding of our universe requires knowledge of the 
dynamics, the quantitative role of gravitational forces determining 
the scale factor evolution, $a(t)$.  This echoes the 
flows of energy between components, e.g.~the epoch of radiation domination 
transitioning to that of matter domination, and is a key element in 
the growth of density perturbations into structure.  Yet until 
recently the literature tended to consider only 
\beq 
H_0=(\dot a/a)_0\qquad;\qquad q_0=-(a\ddot a/\dot a^2)_0\,,
\label{h0q0}\eeq  
the Hubble constant and the deceleration parameter today. 

Now a myriad of cosmological observational tests can probe the 
function $a(t)$ more fully, over much of the age of the universe 
(see Sandage \cite{san88}, Linder \cite{lin89,lin97}, Tegmark 
\cite{teg}).  
All that is required is a probe 
capable, not just in theory but in practice, of observations both 
precise and accurate enough.  A number of promising methods are 
being developed, but this talk concentrates on the most 
advanced, the magnitude-redshift relation of Type Ia supernovae. 

The goal of 
mapping out the recent expansion history of the universe 
has several motivations.  The thermal history of the universe, 
extending back through structure formation, matter-radiation 
decoupling, radiation thermalization, primordial nucleosynthesis, 
etc. has taught us an enormous amount about both cosmology and 
particle physics.  It has spin offs in high energy physics, neutrino 
physics, gravitational physics, nuclear physics, and so on (see, 
e.g., Kolb and Turner \cite{kt}).  The recent expansion history of the 
universe promises similarly fertile ground with the discovery of 
the current acceleration of the expansion of the universe.  This 
involves concepts of the late time role of high energy field theories 
in the form of possible quintessence, scalar-tensor gravitation, higher 
dimension theories, brane worlds, etc.

Looking literally to the future, this accelerated expansion moreover has 
profound implications for the fate of the universe, from the 
viability of string theory \cite{string} to eternal inflation and the heat 
death of the universe \cite{krauss} to ideas on the cyclic 
nature of time \cite{stein}.  The recent expansion history 
offers guidance on the fate of our universe plus physics at the extremes: 
the form of high energy physics, physics at the smallest 
scales and in extra dimensions, physics in the most distant past 
and asymptotic future. 

Section \ref{sec.map} considers the use of supernova observations 
to obtain the 
magnitude-redshift law out to $z\approx2$ and how to relate this to the 
scale factor-time behavior $a(t)$.  Section \ref{sec.snap} presents 
specifics on the proposed Supernova/Acceleration Probe mission \cite{snap} 
and its capabilities.  Different parametrizations of the 
dynamics are investigated in Section \ref{sec.mod}, including 
extensions to nonstandard gravitation that alters the Friedmann equation 
governing the expansion evolution.  Section \ref{sec.constr} 
considers constraints 
from other probes, especially on the age of the universe and the 
Hubble constant.  

The reconstruction of the recent expansion history, \`a la 
Figure \ref{atsnap}, may hopefully soon be a standard feature of 
future textbooks.

\section{Mapping the Expansion History}\label{sec.map}

First we consider the global description: the geometry of the universe 
and the form 
of the spacetime metric that this imposes.  Readers familiar with 
cosmology might wish to skip this didacticism and proceed to 
either Eq.~\ref{metric} or \ref{ta}. 

Precision measurements of the cosmic microwave background 
(CMB) radiation temperature across the entire sky, as well as 
subsidiary experiments such as radio source counts, galaxy 
counts, and background radiation surveys in other wavelength 
bands, indicate our universe is very well modeled by an 
isotropic spacetime.  Redshift surveys such as the 2dF and 
Sloan Digital Sky Survey allow us to begin to construct a 
three dimensional picture of the universe to test homogeneity 
directly.  CMB fluctuation measurements can constrain 
inhomogeneous models as well.  Moreover, we can always fall 
back upon the Cosmological Principle, which provides a 
strong theoretical expectation that the isotropy observed about 
us can be interpreted as about a random location and therefore 
enforces global homogeneity.  Ellis et al.~\cite{ell} have quantified 
the extent to which this could break down and shown that 
formally isotropy about any three points leads to homogeneity. 

A homogeneous and isotropic universe is described within general 
relativity by the Robertson-Walker metric.  This contains only 
two parameters -- a function $a(t)$ and a constant $k$ -- describing 
respectively the scale evolution or dynamics, and the spatial 
curvature.  CMB measurements have further put tight constraints 
on the spatial curvature, restricting its characteristic scale 
today to be of order ten times the horizon scale, i.e.~its 
effective energy density is at most of order 1\% of the total energy 
density. 

Thus for the rest of this paper we adopt the Robertson-Walker 
metric with flat spatial sections, $k=0$: 
\beq
ds^2=-dt^2+a^2(t)[dr^2+r^2(d\theta^2+\sin^2\theta\, d\phi^2)]\,.
\label{metric}\eeq
We will also find it convenient sometimes to use conformal 
time $d\eta=dt/a$. 

Type Ia supernovae, or any standardizable candles (sources with 
known luminosity), are excellently suited to map the expansion history 
$a(t)$ since there exists a direct relation between the observed 
distance-redshift relation $d(z)$ 
and the theoretical $a(t)$.  The scale factor $a$ trivially 
translates into the observable redshift $z$ of the source by 
$a=(1+z)^{-1}$.  That is, the expansion of spacetime directly stretches 
the wavelengths of the emitted light.  On the other hand, that 
light was emitted a finite time ago since it had to traverse 
a certain distance $d$ from the supernova to the observer. 
The received energy flux from the supernova obeys the (relativistic) 
inverse square law, so the distance can be derived from the 
observed flux by relating it to the known emitted flux.  Via 
the speed of light the distance is translated into a ``lookback 
time'' $t$.  Thus one can proceed very simply from observables $d(z)$ 
to the expansion $a(t)$. 

Mathematically, things are almost as simple.  The relation between 
the scale factor and redshift holds for any ``quiet'' universe where 
the light from a source moving at velocities slow 
compared to the Hubble velocity (which is near unity for the 
distant sources used) propagates without interaction in an adiabatically 
evolving background.  The distance used is the luminosity 
distance $d_L$, which is related to the coordinate distance 
appearing in the metric (\ref{metric}) by $d_L=(1+z)r$.  All that remains 
is to find $r(z)$. 

The form of $r(z)$ comes from light following null geodesics, $ds=0$. 
Thus $dr=dt/a=d\eta$ and 
\beq 
r(z)=\eta(z)=\int_{a_e}^1 da/(a^2H)=\int_0^z dz'/H(z')\,,
\label{rh}\eeq 
where the Hubble parameter (a function now, not a single number) is 
$H=\dot a/a$ and $a_e=1/(1+z)$ is the scale factor at the time of 
emission, i.e.~when the supernova exploded.  This is the kinematical 
result, following purely from the geometry.  To evaluate the 
function we need to introduce dynamics: equations of motion derived 
from the gravitation theory.  

In general relativity these are known 
as the Friedmann equations and the relevant one relating the Hubble 
parameter to the matter and energy contents of the flat universe is 
\beq
H^2=(8\pi/3)\rho\,.
\label{hrho}\eeq 
The conservation condition of each (noninteracting) component is 
\beq
\dot\rho/\rho=-3H(1+p/\rho)\equiv -3H[1+w(z)],
\label{rhocons}\eeq 
where the energy density is $\rho$, the pressure $p$, 
and the equation of state (EOS) of each component is defined by 
$w=p/\rho$.  Ordinary nonrelativistic matter has $w=0$; 
a cosmological constant has $w=-1$.  We explicitly allow the 
possibility that $w$ evolves.  The total density and pressure are 
just the sum of the individual components.  

\begin{figure}[!hb] 
\begin{center} 
\psfig{file=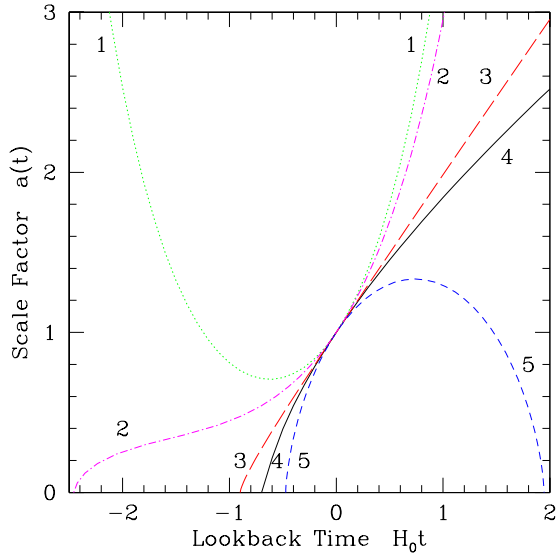,width=3in} 
\caption{
Overall expansion histories of expanding universes are sorted 
into classes: 1) a bounce model, 2) a loitering Eddington-Lema\^\i tre 
model, 3) an open model, 4) a critical, Einstein-de Sitter model, 5) 
a closed model. 
} 
\label{atgross}
\end{center} 
\end{figure}

\begin{figure}[!hb]
\begin{center} 
\psfig{file=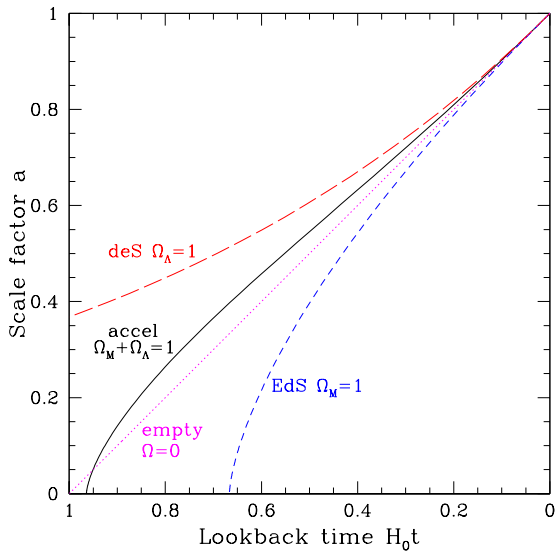,width=3in} 
\caption{A more refined picture of the expansion history shows the 
differences between matter dominated (flat, Einstein-de Sitter), empty 
($\Omega=0$, Milne), cosmological constant dominated (de Sitter), and 
presumably our universe (solid line): an accelerating model with 
both matter and dark energy. 
} 
\label{tasimp}
\end{center} 
\end{figure}

Once one has $H(z)$ one can map out the expansion history $a(t)$ 
by 
\beq
t(a)=\int_a^1 da'/(a'H)=\int_0^z dz'/[(1+z')H(z')].
\label{ta}\eeq 
The gross behavior of the scale factor over the entire age of the 
universe is illustrated in Figure~\ref{atgross}. 
Note that the lookback time is zero at $z=0$ ($a=1$).  One can classify 
expanding cosmologies into open, closed, and critical cases, and 
those which possessed a Big Bang vs.~bounce models.  To set the 
stage by adding one level of detail at a time, we are next interested 
in discrimination between models 
which have less extreme asymptotic behaviors.  A blowup of the past 
history of simple models appears in Figure~\ref{tasimp}. 
The main focus of this paper though is a much finer discrimination, 
between models distinguished by small differences in their components. 
This is probed through the recent expansion history.

\section{SNAP Capabilities}\label{sec.snap}

\begin{figure}[!h]
\psfig{file=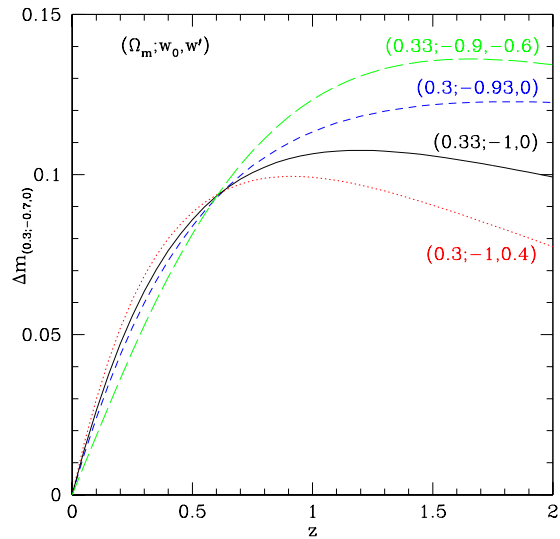,width=3in} 
\caption
{Degeneracies due to the dark energy model and the cosmological model 
cannot be resolved at low redshifts.  In this differential 
magnitude-redshift diagram the three 
parameters to be determined are varied two at a time. Only at $z\approx1.7$ 
do these very different physics models first exceed 0.02 mag 
discrimination. From 
\cite{HL02}.
}
\label{fig:degen}
\end{figure}

The distance-redshift relation can map out the expansion history. 
From the ability of supernovae, or other probes, to observe $r(z)$ 
one can fit and constrain $\rho$ and $w(z)$ of each component. 
To investigate the dark energy and distinguish between classes of 
physics models we need to probe the expansion back into the 
deceleration epoch, 
indeed over a redshift baseline reaching $z>1.5$ (see Fig.~\ref{fig:degen}; 
\cite{Li01,HL02}). 
SNAP \cite{snap} is a simple, dedicated experiment specifically designed to 
map the distance-redshift relation out to $z=1.7$ with high precision 
and tight control of systematic errors. 

These data 
can determine the cosmological parameters with high precision: 
depending on the exact model, the mass density
$\Omega_m$ to $\pm0.01$, vacuum energy density $\Omega_\Lambda$ 
and curvature $\Omega_k$ to $\pm0.03$, and 
the dark energy equation of state $w$ 
to $\pm 0.05$ and its time variation $w'=dw/dz$ to $\pm0.3$.  
This time variation is a crucial distinguishing feature, not only 
for ruling out a cosmological constant explanation, but for guidance 
on the proper class of high energy physics theory to pursue.  In 
addition, wide area weak gravitational lensing studies with SNAP 
will map the distribution of dark matter in the universe and teach 
us about the evolution of the nonlinear mass power spectrum. 

The SNAP mission concept is a 2.0 meter space telescope with a nearly 
one square degree field of view.  A half 
billion pixel, wide field imaging 
system comprises 36 large format new technology CCD's and 36 
HgCdTe infrared detectors.  Both the imager and a low resolution ($R\sim 100$) 
spectrograph cover the wavelength range 3500 - 17000 \AA, allowing 
detailed characterization of Type Ia supernovae out to $z=1.7$. 

As a space experiment SNAP will be able to study supernovae over a much 
larger range of redshifts than has been possible with the current 
ground-based measurements -- over a wide wavelength range unhindered by 
the Earth's atmosphere and with much higher precision and accuracy. 
Many of these systematics-bounding measurements are only achievable 
in a space environment with low sky noise and a very small 
and stable point spread function (critical for lensing as well).  
Unlike other cosmological probes, supernova studies have 
progressed to the point that a detailed catalog of known and possible 
systematic uncertainties has been compiled -- and, more importantly, 
approaches have been developed to constrain each one.  

An array of data (e.g.~supernova 
risetime, early detection to eliminate Malmquist bias, 
lightcurve peak-to-tail ratio, identification of the Type Ia-defining 
Si II spectral feature, separation of supernova light from host galaxy 
light, and identification of host galaxy morphology, etc.) 
makes it possible to study each individual supernova and measure 
enough of its physical properties to recognize deviations from standard 
brightness subtypes.  
For example, an approach to the problem of possible supernova evolution 
uses the rich stream of information that an expanding supernova 
atmosphere sends us in the form of its spectrum.  
A series of measurements will be constructed for each supernova that 
define systematics-bounding subsets of the Type Ia category.  
Only the change in brightness as a function of the 
parameters classifying a subtype is needed, not any intrinsic brightness. 
By matching like to like among the supernova subtypes, we can 
construct independent Hubble diagrams for each, which when compared bound 
systematic uncertainties at the targeted level of 0.02 magnitudes. 

With a prearranged photometric observing program one obtains 
a uniform, standardized, calibrated dataset for each supernova, 
allowing for the first time comprehensive comparisons across complete 
sets of supernovae.  The observing requirements also yield 
data ideal as survey images, and one automatically obtains host galaxy 
luminosity, colors, morphology, and type -- a rich resource 9000 times 
larger than the Hubble Deep Field and somewhat deeper.  Thus SNAP will 
map the distance-redshift relation and much more.  

\section{Modeling the Dynamics}\label{sec.mod}

A fly in the theoretical ointment is that the measured 
distance $r(z)$ is related to, but 
is not, the desired history relation $a(t)$.  So we need to translate 
$r(z)$ into $a(t)$; this requires an intermediate step of obtaining 
$H(z)$.  We can do this either directly or through the cosmology 
parameters $\rho$ and $w(z)$.  The direct method involves a 
derivative of $r(z)$, so noisy data can introduce difficulties 
\cite{astier,ht}.  We examine this approach further in 
\S\ref{sec.hdirect}.  First we discuss the reconstruction of $H(z)$ 
from the fit of the cosmological parameters to the observations. 

Observational evidence for accelerated 
expansion informs us that (within the dark energy picture; 
cf.~\S\ref{sec.bey}) there must 
be further cosmological parameters, describing another component with 
a strongly negative EOS. 

Assuming that just these two components, matter and another with EOS 
$w(z)$, control the dynamics during the epoch of interest, we obtain 
a solution for $H(z)$ and hence $r(z)$ 
by combining equations (\ref{rh})-(\ref{rhocons}): 
\begin{eqnarray}
H_0r(z) & = & \int_0^z dz'\,\Big[\om(1+z')^3 
+(1-\om) \nonumber\\ 
& \times & e^{3\int_0^{\ln(1+z')} 
d\ln(1+z'') [1+w(z'')]}\Big]^{-1/2}, 
\label{rwz}
\end{eqnarray}
where $\om$ is the dimensionless matter density $8\pi\rho_m/(3H_0^2)$ 
and $H_0$ is the Hubble constant -- the present value of the Hubble 
parameter.  Equation (\ref{ta}) can then be used to obtain $a(t)$. 

The EOS $w(z)$ is derived from the Lagrangian for that component, 
i.e.~the particle physics enters here.  One could solve the scalar field 
equation for a particular model to find $w(z)$ but then one does not 
obtain a model independent parameter space in which to compare models. 
For generality of treatment, 
various parametrizations of $w(z)$ are used (though \cite{hutstar} 
introduces a principal component approach).  We discuss a standard 
form next and a new parametrization in 
\S\ref{sec.newpar}.  

\subsection{Linear $w(z)$}\label{sec.linpar}

\begin{figure}[!hb]
\begin{center} 
\psfig{file=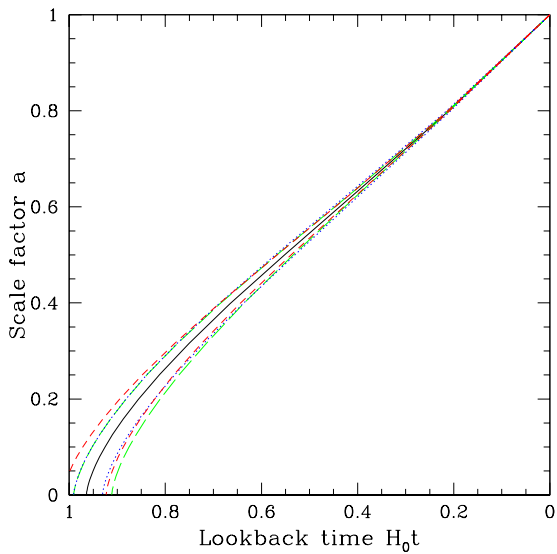,width=3in} 
\caption{Precision experiments are necessary to differentiate the 
expansion histories of models with different densities and equations 
of state.  The central curve is the history of a flat $\Omega_m=0.3$ 
model with cosmological constant.  The red dashed curves vary the matter 
density by $\pm0.05$, the blue dotted ones vary the equation of state 
of dark energy by $\pm0.2$, and the green long dashed curves put in 
a time variation of the dark energy EOS, $dw/dz=\pm0.5$. 
} 
\label{atdw}
\end{center} 
\end{figure}

The conventional first order expansion to the EOS, enlarging the 
phase space to 
incorporate the critical property of time variation in the EOS, 
is $w(z)=w_0+w_1z$.  In this case the 
exponential in (\ref{rwz}) resolves to $(1+z)^{3(1+w_0-w_1)}e^{3w_1z}$. 
Figure~\ref{atdw} illustrates 
the effect of changing the cosmological parameters, one at a time, 
on the expansion history. 

\begin{figure}[!h]
\begin{center} 
\psfig{file=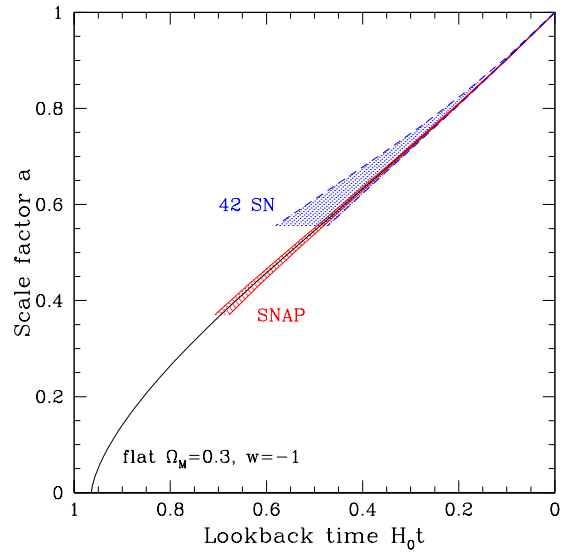,width=3in} 
\caption{The constraints that supernova mapping of the expansion   
history through the magnitude-redshift relation provide.  SNAP limits 
represent a generational advance in understanding the recent history of 
the universe. 
} 
\label{atsnap}
\end{center} 
\end{figure}

An important point is the 
presence of correlation between the parameters $\om$, $w_0$, 
$w_1$ (see, e.g., \cite{welal}) which must be treated properly 
if more than one is 
allowed to vary (which is of course the general case).  
Figure~\ref{atsnap} shows the 
reconstructed expansion history for a simulation of the 
future SNAP experiment. Despite 
it being able to determine each parameter individually to high precision, 
e.g.~$\om$ to 0.03, $w_0$ to 0.05, $w_1$ to 0.3 (each marginalized 
over others), the correlations 
among them (i.e.~degeneracies among their combinations) 
relax the tightness of the constraint SNAP would 
place on the expansion history.  This is unavoidable (but see 
\S\ref{sec.conformal}).  

\subsection{A new parametrization of the dark energy}\label{sec.newpar}

Mapping the expansion history out to redshifts $z>1$, 
beyond the deceler\-ation-acceleration transition 
represents a major advance in our cosmological knowledge.  
But high redshift does introduce 
complications in the parametrization of dark energy.  In 
order to draw model independent constraints, we had 
parametrized the dark energy EOS linearly in redshift: 
$w(z)=w_0+w_1z$.  This clearly grows increasingly problematic 
at redshifts $z>1$. However exact solutions for the EOS from 
the scalar field equations of motion do not allow us to make 
model independent statements. 

Corasaniti and Copeland \cite{cc} originally suggested an alternate 
parametrization in terms of a 7-dimensional phase space: 
the value of the EOS today, $w_0$, the values deep in the matter 
(radiation) dominated tracker regime, $w_m$ ($w_r$), the scale 
factors $a_{cm}$ ($a_{cr}$) at the time the field leaves the tracking 
behaviors, and the widths $\dl_m$ ($\dl_r$) of those transitions 
in Hubble units.  They find success 
in a particular functional approximation with scale factor, 
obtaining the EOS $w(z)$ to better than 5\% back to the last 
scattering surface for a range of models that display tracker 
behavior (where the field is in a slow roll regime that 
keeps the EOS nearly constant, determined by the EOS of the 
background component).  Indeed they extend this function back 
into the radiation dominated epoch as well.  While this only 
applies to dark energy field theories that have a slow roll 
regime, within that fairly varied class (e.g.~potentials with 
inverse power laws, double exponentials, supergravity 
inspired models, etc.) it provides a model independent method 
of characterizing the behavior. 

Such generality, while impressive, is somewhat unwieldy.  For our present 
purposes we can make two simplifications.  First, since we aim 
only to trace the expansion history back to the matter 
dominated epoch (for now at least), I simplify the phase space to 
four dimensions and obtain a fitting function for $w(a)$ as  
\beqa
w(a)&=&F+G\,\left[1+e^{-(a-a_c)/\dl}\right]^{-1} \\ 
F&=&w_0-(w_m-w_0)(1+e^{-a_c/\dl})(e^{-1/\dl}-1)^{-1} \\ 
G&=&(w_m-w_0)(1+e^{-a_c/\dl})(e^{-1/\dl}-1)^{-1} \nonumber\\ 
& \times & (1+e^{-(1-a_c)/\dl})\,. 
\label{wa4d}\eeqa 
Out to the last scattering surface at 
$z=1100$ this is as accurate as the original Corasaniti \& 
Copeland expression.  The time variation of the EOS, evaluated 
at $z=0$, is 
\beqa 
w' & \equiv & (dw/dz)_0=(w_m-w_0)\dl^{-1}(e^{1/\dl}-1)^{-1} \nonumber\\  
& \times & (1+e^{a_c/\dl})(1+e^{-(1-a_c)/\dl})^{-1}. 
\label{w1cc}\eeqa 

But analyzing the constraints of SNAP or other cosmological 
probes on a 4-dimensional dark energy parameter space, in 
addition to other parameters such as $\om$ and the supernova 
intrinsic magnitude, is too broad for useful conclusions. 
Instead I make a second simplification by using a fitting 
function 
\beqa
w(a)&=&w_0+w_a(1-a) \\ 
&=&w_0+w_a z/(1+z). 
\label{wa}\eeqa 
This is astonishingly successful (see Fig.~1 in \cite{lin0210}). 

This new parametrization\footnote{A few months after this talk, 
D.~Polarski kindly directed me to \cite{polwa}, though there they 
do not consider $z\gg1$.} of dark energy models has several 
advantages: 1) a manageable 
2-dimen\-sional phase space, the same size as the old $w_0-w_1$ 
parametrization, 2) reduction to the old linear redshift 
behavior at low redshift, 3) well behaved, bounded behavior 
for high redshift, 4) high accuracy in reconstructing many 
scalar field equations of state and the resulting distance-redshift 
relations, 5) good sensitivity to observational data, 6) simple 
physical interpretation.  Particularly 
important is its virtue of keeping $w(a)$ of order unity even 
for large redshifts; this is essential when analyzing cosmic 
microwave background (CMB) constraints (to $z=1100$) on dark 
energy.  This contrasts with the linear redshift expansion 
from the previous section. 

Beyond the bounded behavior, though, the new parametrization is 
also more accurate than the old one.  For example, in comparison 
to the exact solution for the supergravity inspired SUGRA model 
\cite{braxm} it is accurate in 
matching $w(z)$ to -2\%, 3\% at $z=0.5$, 1.7 vs.~6\%, -27\% for 
the linear $z$ approximation (the constants $w_1$, $w_a$ 
are here chosen to fit at $z=1$).  Most remarkably, 
it reconstructs the distance-redshift behavior of the SUGRA 
model to 0.2\% over the entire range out to the last scattering 
surface ($z\approx1100$). The physical interpretation of the 
parametrization is straightforward: 
$w_0$ is the present value of the EOS and $w_a$ is a measure 
of the time variation, which can be chosen to give the correct 
value of $w(z=1)$.  For the cosmological constant, of course 
$w_a=0$.   

Figure~\ref{atdw} shows lines of constant 
$w_a=\pm1$ ($dw/dz\approx0.5$) in the expansion evolution $a(t)$.  
Note that for 
$w(a)=w_0+w_a(1-a)$ the dark energy 
density exponential in (\ref{rwz}) resolves to $a^{-3(1+w_0+w_a)} 
e^{-3w_a(1-a)}$. 

Also note that $dw/d\ln(1+z)|_{z=1}=w_a/2$; one might 
consider this quantity a natural measure of time variation (it is 
directly related to the scalar field potential slow roll factor $V'/V$) 
and $z=1$ 
a region where the scalar field is most likely to be evolving as the 
epoch of matter domination begins to change over to dark energy 
domination. 

SNAP will be able to determine $w_a$ to better than $\pm0.55$ (one 
expects roughly $w_a\approx 2w_1$), with use of a prior on $\om$ of 
0.03, or to better than 0.3 on incorporating data from the Planck CMB 
experiment \cite{planck}. For the advantages of combining supernova 
and CMB data see \cite{fhlt}.  The CMB information can be folded in 
naturally in this parametrization, without imposing artificial 
cutoffs or locally 
approximating the likelihood surface (Fisher matrix approach) 
as required for SNe plus a 
CMB prior in the $w_1$ parametrization. In fact, the new 
parametrization is even more promising since the sensitivity of the 
SNAP determinations increases for $w_0$ more positive than $-1$ 
or for positive $w_a$ (see, e.g., \cite{astier}): the values quoted 
above were for a fiducial cosmological constant model.  For example, 
SUGRA predicts $w_a=0.58$ and SNAP would put error bars of 
$\sigma(w_a)\approx0.25$ on that; this would demonstrate time 
variation of the EOS at the 95\% confidence level.  
Incorporation of a Planck prior can improve this to 
$\sigma(dw/d\ln(1+z)|_{z=1})\approx0.1$, i.e.~the $\approx$99\% 
confidence level.  

\subsection{Using $H(z)$ directly}\label{sec.hdirect}

The direct reconstruction method of going from observations 
$r(z)$ to $H(z)$, and then to $a(t)$, without a parametrization 
in terms of $w(z)$ can be attempted.  This has the virtue of 
model semi-independence, allowing incorporation of additional 
errors, e.g.~systematics, outside the distance relation.  
But since it can only be 
carried out in a local perturbative manner (\`a la the Fisher 
matrix) it does depend on a known fiducial model.  But this 
is a model merely for $H(z)$, not for the details $w(z)$, 
i.e.~it is a nonparametric reconstruction, and it can give a 
feel for the effect of measurement errors. 

Here we relate the derived uncertainties about a fiducial 
behavior $H(z)$ to the observational errors. 
First we must realize 
that the supernovae observations are phrased in terms of 
magnitudes, or logarithmic fluxes: $m(z)\sim 5\log 
[(1+z)\,r(z)]$, so one really is interested in $\sigma_{H(z)}$ as 
a function of the measurement errors $\sigma_{m(z)}$.  However, 
a one to one mapping between these quantities at a single 
redshift does not exist, since $m(z)$ involves an integral over the 
Hubble parameter (see Eq.~\ref{rh}).  Instead, a variational 
calculation yields 
\beq 
\sigma_{H(z)}=H(z){\ln10\over5}\left[\sigma_{m(z)}+{d\sigma_m\over dz} 
H\eta\right], 
\label{sighm}\eeq 
i.e.~the parameter error involves both the magnitude error at that 
redshift and its derivative.  Recall $\eta(z)=\int_0^z dz'/H$ is 
the comoving distance or conformal time. 

For a fiducial model $H(z)$ one can translate errors in the 
magnitude data into uncertainties on the Hubble parameter, 
and then further to the $a(t)$ relation.  This is convenient 
for taking into account the realistic situation of not only 
uncertainty in the cosmology parameters but in the observational 
data.  
For example, one can analyze the effect of statistical 
and systematic magnitude errors on the mapping of the 
expansion history.  To carry the error progragation one step 
further, to $a(t)$, we must integrate the Hubble parameter we found from  
differentiating the data in order to obtain the lookback time.  The 
resulting error in the time $t(a)$ is now nonlocal in redshift:  
\beqa 
\sigma_{t(z)} & = & \sigma_{m(z)}{\ln10\over5}(1+z)^{-1}\eta \nonumber\\ 
& + & {\ln10\over5} 
\int_0^z dz'\,(1+z')^{-2}\eta(z')\,\sigma_{m(z')} \,.
\eeqa  
In the next section we will see a better solution. 

\subsection{Conformal Time History}\label{sec.conformal}

One method of incorporating the advantages of both approaches -- 
the generality of parametrization and the directness of 
reconstruction -- is to 
alter slightly our view of the expansion evolution.  Instead 
of $a(t)$, consider the conformal time $a(\eta)$.  From 
$d=(1+z)\eta$ one sees that one requires no foreknowledge 
or local approximation to obtain the scale factor-conformal 
time relation. 

We have 
\beqa 
m(z)&\sim &5\log[(1+z)\eta(z)],\\ 
\eta(a)&\sim&a\cdot 10^{m(z)/5}=\eta_0(a)\cdot 10^{dm(z)/5}
\label{maeta}\eeqa 
where in the last equality we explicitly show how the conformal 
time-scale factor relation changes in the presence of a shift 
or errors in the observed magnitude.  This error propagation then 
reduces simply to $\sigma_{\eta(z)}=\sigma_{m(z)}\,(\ln10/5)\eta$.  
Thus the fractional uncertainty is 
\beq 
{\sigma_\eta\over\eta}=\sigma_m{\ln10\over 5}\approx(1/2)\sigma_m. 
\label{sigeta}\eeq 
All quantities are at a single redshift.  Only in the case where 
$\sigma_m$ is constant in redshift does one obtain similar 
precision on the other mapping parameters: then $\sigma_t/t=\sigma_H/H=
\sigma_\eta/\eta$.  

\begin{figure}[!hb]
\begin{center} 
\psfig{file=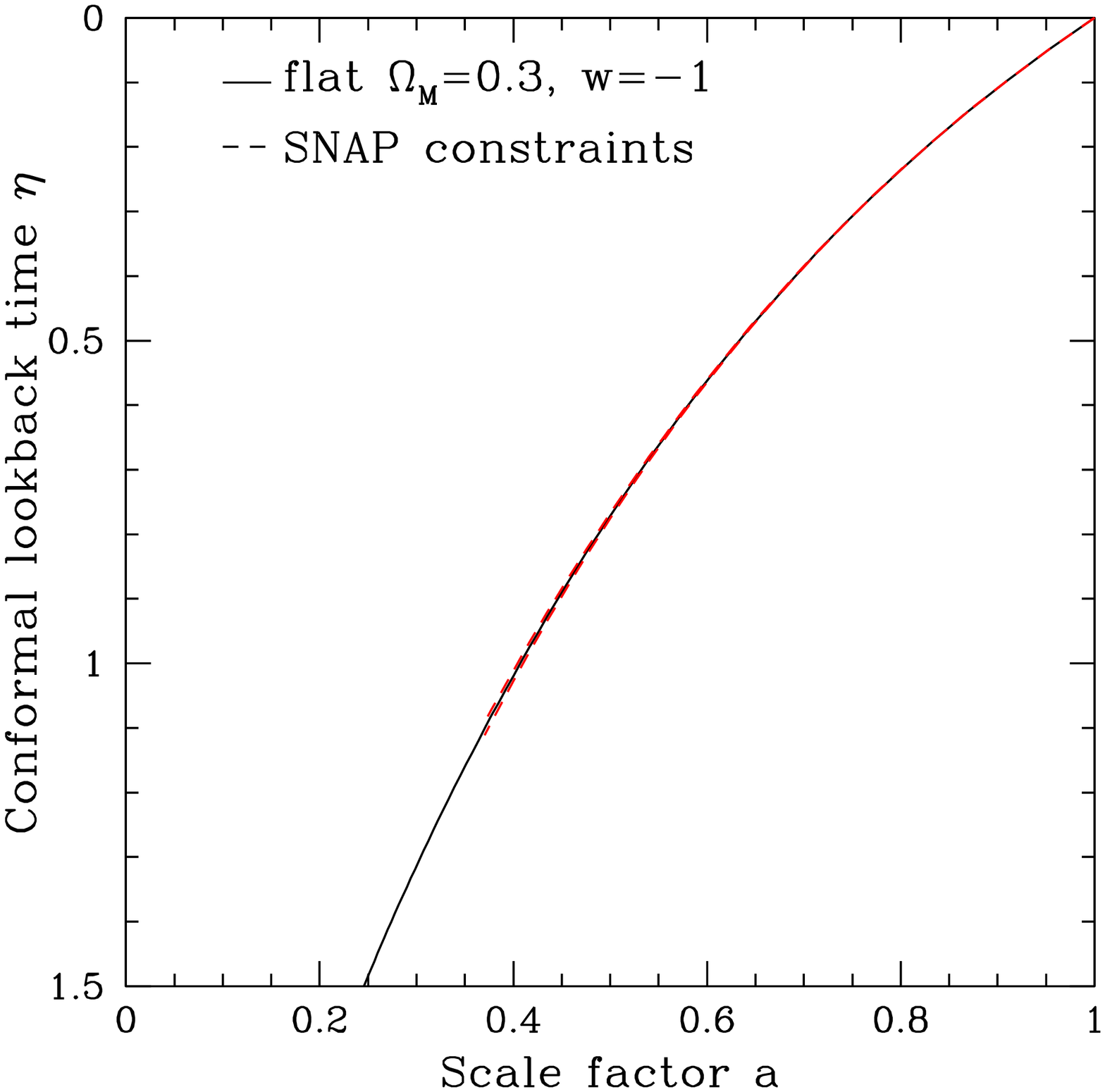,width=3.0in} 
\psfig{file=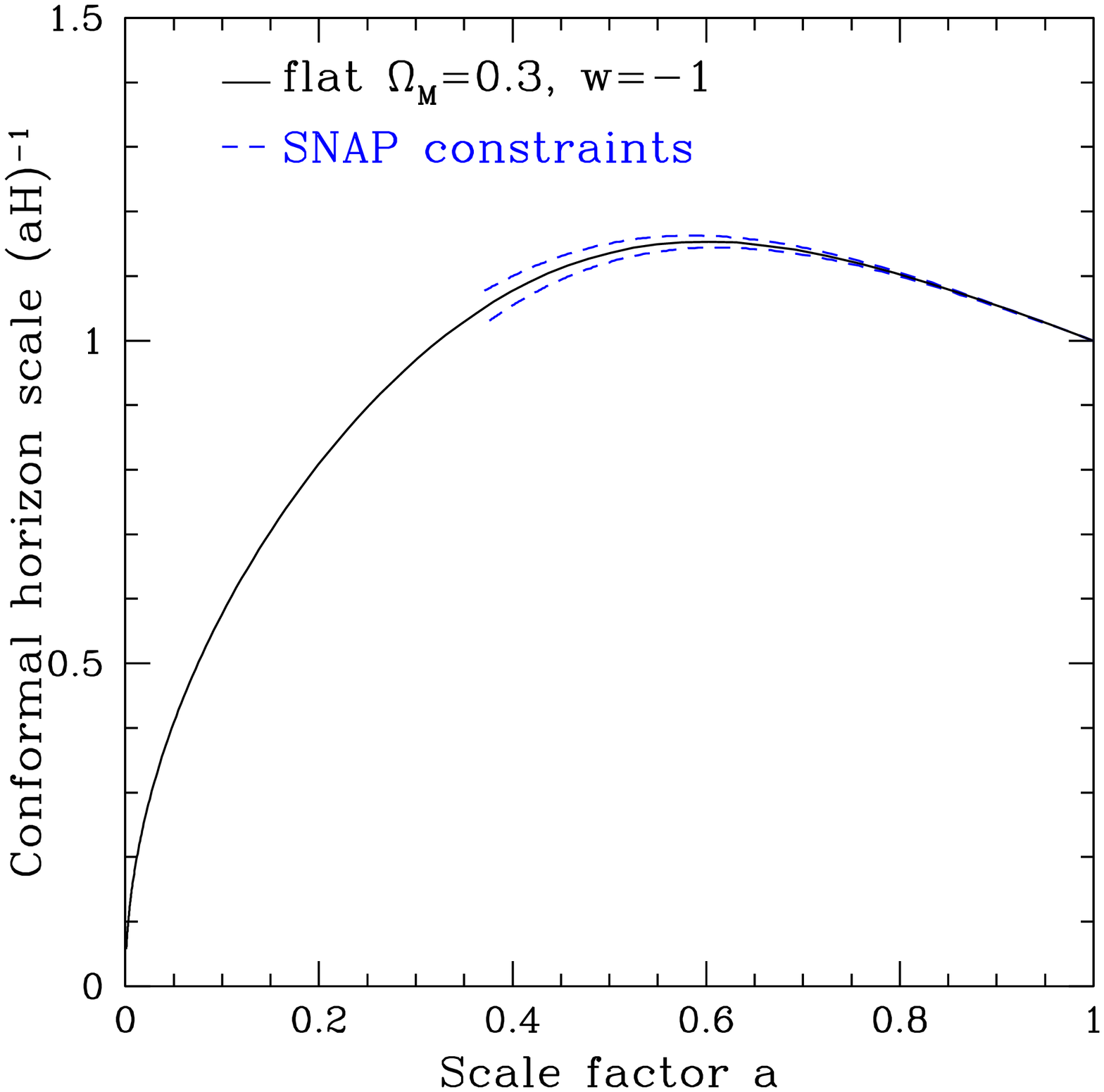,width=3.0in} 
\caption{The expansion history is plotted in conformal time in the 
left figure.  SNAP constraints appear very tight when viewed 
in conformal time because this is most closely related to the 
observations. The right figure shows the 
conformal horizon scale -- the logarithmic derivative of the left 
figure.  The part with negative slope 
allows comoving wavelengths to expand outside the horizon, or alternately 
represents $aH=\dot a$ increasing, i.e.~$\ddot a>0$ -- precisely 
the signature of inflation or acceleration.  
The dashed blue lines show that SNAP will map the accelerating 
phase, the transition, and into the matter dominated, decelerating phase 
of the past universe. 
} 
\label{aeta}
\end{center} 
\end{figure}

Mapping of the expansion 
history in conformal time is shown in Figure~\ref{aeta}.  Note 
that the reconstruction is much tighter due to the 
straightforward translation from observations.  The 1\% distance 
measurement error ($\sigma_m=0.02$) given by SNAP's limiting 
systematics becomes 
a 1\% error in $a(\eta)$.  Moreover there is no need for  
the indirect mapping method of \S\ref{sec.linpar}, which led 
to a $\sim$2\% error in $a(t)$ due to the parameter correlations 
mentioned in that section. 

Figure~\ref{aeta} also shows the 
logarithmic derivative 
\beq 
{d\eta\over d\ln a}=a{dt\over a\,da}=(\dot a)^{-1}=(aH)^{-1},
\eeq 
interpreted respectively as the 
proper time evolution of the scale factor or 
the conformal horizon scale $(aH)^{-1}$.  Just as in inflation 
a positive slope denotes decelerating expansion, 
$\ddot a<0$ and comoving wavelengths (e.g.~of density perturbations) 
enter the horizon; v.v.~for a negative slope: an accelerating universe 
or inflation.  The dashed lines show that SNAP can probe both epochs 
and map the transition between them.

\subsection{Beyond Dark Energy}\label{sec.bey}

Mapping the physical time evolution of the scale factor relies on 
translating the observations into the behavior of the Hubble 
parameter $H(z)$.  This translation might proceed via a 
parametrization of the physics of the acceleration, e.g.~the 
dark energy properties as in \S\ref{sec.mod}.  We also might want to 
study the density evolution and its transition from the earlier 
matter dominated epoch to the present \cite{linexp}. 
In \S2 we adopted general relativity to obtain the Friedmann 
equation (\ref{hrho}) to provide the foundation for these. 

But ideally we would like to use the data to test the 
Friedmann equations of general relativity or alternate explanations 
for the acceleration besides dark energy.  The supernova 
distance-redshift data allow such investigation of the fundamental 
framework by substituting the altered Hubble evolution $H(z)$ into 
Eq.~(\ref{rh}).  This enables constraints to be placed on, e.g., 
higher dimension theories, Chaplygin gas, etc. See Linder \cite{linexp} 
for examples.

\section{Complementary Probes}\label{sec.constr}

In addition to the mapping of the expansion history 
through the distance-redshift relation by Type Ia supernovae or possibly 
other methods in the future, one can constrain the $a(t)$ curve in 
other ways.  The total age of the universe places the ``foot'' of the 
$a(t)$ curve on the $a=0$ axis in the lower left of the plots.  
``Shooting upward'' with a known slope ($a\sim t^{1/2}$ in the early 
radiation dominated epoch) provides a constraint on the $a(t)$ relation. 
Knox et al.~\cite{knox} have shown that the location 
of the acoustic peaks in the cosmic microwave background radiation 
power spectrum is nearly degenerate with the age in a flat universe 
not too different from ours. 
These data place a constraint of $t_0=14.0\pm0.5$ billion years.  The 
bounds in terms of $H_0t_0$ are slightly weaker because of increased 
model dependence: $0.93\pm0.06$, $1.00\pm0.07$ from supernovae 
\cite{saul,h0t0}. 

The $t$, as opposed to the $H_0t$, axis would thus have a 
stronger footpoint for the $a(t)$ curve to aim toward.  But conversely, 
the slope of $a(t)$ near $a=1$ changes from unity on a $H_0t$ plot 
to the Hubble constant $H_0$ on a $t$ plot, adding 
another uncertainty.  A direct bound on $H_0$ 
constrains the slope of the $a(t)$ curve at $a=1$ in the upper right 
of such plots.  The Hubble Space Telescope Key Project \cite{freed} 
quotes $H_0=72\pm8$ km/s/Mpc.  In either the $a-t$ or $a-H_0t$ plane 
constraints on present slope and total age together act to 
force the expansion history into a narrow corridor, shooting back 
in time subject to the boundary conditions. 

\section{Conclusion} 

The geometry, dynamics, and composition of the universe are intertwined 
through the theory of gravitation governing the expansion of the universe. 
By precision mapping of the recent expansion history we can hope to 
learn about all of these.  The brightest hope for this in the near 
future is the next generation of distance-redshift measurements through 
Type Ia supernovae that will reach out to $z\approx1.7$.  This 
represents over 70\% of the age of the universe and spans the 
current accleration 
epoch back to the matter dominated deceleration epoch when most 
large scale structure formed. 

Just as the thermal history of the early universe taught us much about 
cosmology, astrophysics, and particle physics, so does the recent 
expansion history have the potential to greatly extend our physical 
understanding.  With the new parametrization of dark energy suggested 
here, one can study the effects of a time varying equation of state 
component back to the decoupling epoch of the cosmic microwave background 
radiation.  But even beyond dark energy, exploring the expansion history 
provides us cosmological information in a model independent way, allowing 
us to examine many new physical ideas.  From two numbers we have 
progressed to mapping the entire dynamical function $a(t)$, on the 
brink of a deeper understanding of the dynamics of the universe. 

\section*{Acknowledgments}

This work was supported at LBL by the Director, Office of Science, 
DOE under DE-AC03-76SF00098.  I thank Alex Kim, Saul Perlmutter, and 
George Smoot 
for stimulating discussions, Ramon Miquel and Nick Mostek for 
stimulating computations, and of course Jos\'e Nieves for organizing 
such a stimulating conference.

\end{document}